\documentclass[12pt,preprint]{aastex}
\slugcomment{Accepted to ApJ}

\shorttitle{Periodic Oscillations in S5 0716$+$714}
\shortauthors{Gupta et al.}

\begin{document}


%

\title{Periodic Oscillations in the Intra-day Optical Light Curves 
of the Blazar S5 0716+714}


\author{Alok C. Gupta\altaffilmark{1}, A. K. Srivastava\altaffilmark{1}, 
and Paul J.\ Wiita\altaffilmark{2,3}}

\altaffiltext{1}{Aryabhatta Research Institute of Observational Sciences (ARIES),
Manora Peak, \\
\hspace*{0.22in} Nainital $-$ 263129, India}
\altaffiltext{2}{Department of Physics and Astronomy, Georgia State University, Atlanta, 
GA 30302--4106}
\altaffiltext{3}{School of Natural Sciences, Institute for Advanced Study,
Princeton, NJ 08540}

\email{acgupta30@gmail.com, aks@aries.ernet.in, wiita@chara.gsu.edu \\
Phone No. +91-9936683176, Fax No. +91-5942-233439}
%

%
%

\begin{abstract}

We present results of a periodicity search of 20 intra-day variable 
optical light curves of the blazar S5 0716$+$714, selected from
a database of 102 light curves spanning over three years. We use a 
wavelet analysis technique along with a randomization test and 
find strong candidates for nearly periodic
variations in  eight light curves, with probabilities ranging from 
95\% to $>$99\%. This is the first good evidence 
for periodic, or more-precisely, quasi-periodic, components in the 
optical intra-day variable light curves of any blazar. 
Such periodic flux changes support the idea that 
some active galactic nuclei variability, even in blazars, is based on accretion 
disk fluctuations or oscillations. These intra-day variability time scales are 
used to estimate that the central black hole of the blazar S5 0716$+$714 
has a mass $> 2.5 \times 10^6$ M$_{\odot}$. As we did not find any correlations 
between the flux levels and intra-day variability time scales, it appears that 
more than one emission mechanism is at work in this blazar.

\end{abstract}






\section{Introduction}

Blazars are the subclass of radio-loud active galactic nuclei (AGNs) consisting 
of BL Lacertae objects (BL Lacs) and flat spectrum radio quasars (FSRQs). 
BL Lacs show largely featureless optical continua.  All blazars exhibit strong 
flux variability on diverse time scales varying from a few minutes to many years at 
all accessible wavelengths of the electromagnetic spectrum. For convenience, 
blazar variability can be broadly divided into three classes, viz., intra-day 
or intra-night variability, short-term variability and long-term 
variability. Variations in flux of up to a few tenths of a magnitude over 
the course of a night or less is variously known as intra-day variability (IDV) 
(Wagner \& Witzel 1995), which is the term we adopt here, microvariability 
(e.g., Miller, Carini \& Goodrich 1989), or intra-night optical
variability (e.g., Gopal-Krishna, Sagar \& Wiita 1995) . Short- and 
long-term variabilities can amount to four or five magnitudes and are
usually defined to have time scales from weeks to several months and 
several months to years, respectively. 

It is widely accepted that the central engines of AGNs fundamentally 
are comprised of super massive black holes (SMBHs) along with the radiating 
matter accreting onto them.  In the case of radio-loud AGN such as blazars, 
jets emerge from these central engines. IDV is most likely produced in close 
proximity to the SMBH (e.g., Miller et al.\ 1989), although an origin further 
away within a turbulent jet (e.g., Marscher, Gear \& Travis 1992) 
is also possible. Minimum IDV time scales of blazars can be used to place 
constraints on sizes of the emitting regions, and, if the radiation is
indeed emitted close to the center, on the masses of the SMBHs. 
Detection of periodic or quasi-periodic oscillations in optical IDV light 
curves of blazars would be strong evidence for the presence of a 
single dominant orbiting hot-spot on accretion disk, or accretion disk 
pulsation (e.g., Chakrabarti \& Wiita 1993; Mangalam \& Wiita 1993). 
Hence the search for periodic or quasi-periodic oscillations in the IDV light 
curves of blazars is of great interest; however, such variations have
not yet been well established. The motivation of the present research program 
is to see if such periodic IDV patterns do indeed exist in the optical light 
curves of blazars.

For the present work, we extracted data from the published literature 
for optical IDV of the BL Lac object S5 0716$+$714. It is one of the 
brightest and therefore most well studied BL Lacs with respect 
to optical IDV. The optical continuum of this blazar is so featureless 
that many attempts made to determine its redshift have failed. 
The non-detection of its host galaxy allowed a lower limit to its 
redshift, $z > 0.3$ (Wagner et al.\ 1996), to be set and 
a later non-detection led to a claim that $z > 0.52$ 
(Sbarufatti, Treves \& Falomo 2005).  However, there has been a very recent 
claim of a host galaxy detection which produces a ``standard candle''
value of $z = 0.31 \pm 0.08$ (Nilsson et al.\ 2008). The duty 
cycle of the source is essentially unity, which implies that the 
source is always in at least a moderately active state (Wagner \& Witzel 1995). 
In the search for optical variability on diverse time scales 
in this source, it has been observed intensely over at least the 
last 15 years (e.g.\ Wagner \& Witzel 1995; Wagner et al.\ 1996; 
Heidt \& Wagner 1996; Ghisellini et al.\ 1997; Sagar et al.\ 
1999; Villata et al.\ 2000; Qian, Tau \& Fan 2002; 
Raiteri et al.\ 2003; Wu et al.\ 2005, 2007; 
Nesci et al.\ 2005; Stalin et al.\ 2006; Gu et al.\ 2006;
Montagni et al.\ 2006; Ostorero et al.\ 2006; Foschini et al.\ 2006; 
Villata et al.\ 2008; Gupta et al.\ 2008 and references 
therein). Five major optical outbursts were reported for this source;
they occurred at the beginning of 1995, in late 1997, in the fall of 2001, 
in March 2004 and in the beginning of 2007 (Raiteri et al.\ 2003; 
Foschini et al.\ 2006; Gupta et al.\ 2008). These 
five outbursts indicate a timescale of long-term variability of 
$\sim 3.0 \pm 0.3$ years (e.g., Raiteri et al.\ 2003).  
Correlated radio/optical short-term variability measured over a month
seemed to reveal some quasi-periodicities for this blazar   
(Quirrenbach et al.\ 1991; Wagner et al.\ 1996).

This paper is structured as follows. In \S 2 we discuss the source of the data
we use and we describe the criteria employed to select the  light curves for 
this study. In \S 3 we describe the wavelet analysis and randomization 
technique we employ in our analysis. Our results are presented in \S 4  and we 
discuss them and draw conclusions in \S 5.

\section{Data and Selection Criteria}

There have been several attempts to search for optical IDV in the blazar 
S5 0716$+$714 over the last decade or so, and a few of the most comprehensive 
studies are by Sagar et al.\ (1999); Wu et al.\ (2005, 2007); 
Montagni et al.\ (2006); Stalin et al.\ (2006); Pollock et al.\ (2007); 
and Gupta et al.\ (2008). While examining these papers, we found that the 
best quality data, in the sense of having a combination of  the longest 
nightly durations, the best time resolution, and nearly uniform sampling, 
was published by Montagni et al.\ (2006). 

We downloaded the Montagni et al.\ (2006) data on 0716$+$714  
taken on 102 nights between 1999 and 2003 from the SIMBAD 
astronomical database\footnote{http://simbad.u-strasbf.fr/simbad/}.
We visually inspected all of these light curves (LCs) to decide which
might be most suitable for a periodicity search.

First we rejected the 37 LCs that had observational gaps during that night.  
Next, the 5 LCs that had median errors more than 0.015 mag.\ were rejected so 
as to obtain a high quality data set. 
For the remaining 60 LCs we obtained the intra-day variability amplitude, $A$, 
defined as (Heidt \& Wagner 1996)
\begin{eqnarray}
A = 100 \times \sqrt {(A_{max} - A_{min})^{2} - 2\sigma^{2}} \hspace*{0.1in}\%,
\end{eqnarray}
where $A_{max}$ and $A_{min}$ are the maximum and minimum magnitudes in the
calibrated LC of the blazar and $\sigma$ is the averaged measurement error 
of that blazar LC. So as to obtain LCs for which the presence of IDV is
unquestionable,  we only tentatively accepted the 30 LCs with $A > 10$\%.  
Finally, we rejected the 10 LCs with data extending over less than 7.5 hours. 

This entire set of selection criteria were satisfied on 20 nights for which the
LCs extend from 7.7 to 12.3 hours.  The dates, filters used, LC durations
and variability amplitudes
for these selected datasets are given in the first four columns of Table 1.

\section{Wavelet Analysis and Randomization Technique}

We use a wavelet plus randomization technique, which has certain advantages
over the commonly used periodogram and Fourier power spectra approaches in 
searching for statistically significant real periodicities in IDV LCs. 
The wavelet analysis computer code ``randomlet" was developed by E.\ O'Shea 
in the environment of IDL (Interactive Data Language) software 
(O'Shea et al.\ 2001). Using this code, we can find statistically significant 
real periodicities (if they exist) in time series data. The program incorporates 
a randomization test as an additional feature to the standard wavelet 
analysis code of Torrence \& Compo (1998) which allows robust non-parametric
estimates of the probabilities that periodic components contribute to
a signal (e.g., Bradley 1968). 

Details of the randomization technique used to examine the existence of 
statistically significant real oscillation periods were given by 
Linnell Nemec \& Nemec (1985) for investigations of stellar 
pulsation periods and by O'Shea et al.\ (2001) in the context of solar
coronal variability. This technique has led to 
several important results in the context of solar physics
 by analyzing approximately uniformly sampled 
data (e.g., Banerjee et al.\ 2001; O'Shea et al.\ 2001, 2005; 
Ugarte-Urra et al.\ 2004; Popescu et al.\ 2005; Srivastava et al.\ 2008a,b). 
A similar technique was used by Mathioudakis et al.\ (2006) in 
searching for periodic variations in the active star EQ PegB. 

Here we use this tool for the first time to search for 
periodicity in blazar optical IDV data that 
are also approximately uniformly sampled.  We note that 
there have been previous
efforts to study other types of  blazar variability using variants
of the wavelet technique.  For example, Kelly et al.\ (2003) used a 
continuous wavelet transform
and cross-wavelet  analysis to search for long time scale 
variations in the multi-band radio LCs of 30 AGN; 
they found a quasi-periodic variability in at least one of the radio 
frequencies at which these AGN were monitored for a majority of them.
Examination of the radio variability of PKS B0048$-$097 with the same
technique found a quasi-period of $\sim$ 460 days that later was 
replaced by a more
precise period of $\sim$ 585 days  (Kadler et al.\ 2006).
A wavelet analysis of 19 X-ray LCs from 10 AGNs recently led to
the claim of the presence of a 3.3 ks quasi-period in one data train  
from the quasar 3C 273 (Espaillat et al.\ 2008).  
The ``randomlet'' approach gives
an easily understood way to employ wavelets along with a well-established
statistical test for the reality of 
claimed periodicities in the time series data of the blazar 0716$+$714. 

In all wavelet analyses, the search for periodicities in light curves
 is carried out through a time localized function that is continuous 
in both frequency and time. The
wavelet used in this study is the Morlet function which is defined as
\begin{equation}
\psi_{t} (s) = \pi^{-1/4} \ {\rm exp}(i \omega t) \ {\rm exp} \left (\frac {-t^{2}}{2s^{2}} \right)~,
\end{equation}
where $t$, $s$, $\omega$ and $\pi^{-1/4}$ are the time parameter, 
wavelet scale, oscillation frequency parameter, and the normalization constant, respectively. 
A Morlet function involves the product of a sine wave with a
Gaussian envelope.
The non-dimensional oscillation frequency parameter ($\omega$) 
is set equal to  6 in order to satisfy
the admissibility condition (Farge 1992). 
The Fourier period $P$ is related to the wavelet scale $s$ in the 
Morlet function by the simple relation, $P = 1.03 s$.

The wavelet is convolved with the time series to determine the 
contribution of that frequency 
to the time series through matching the sinusoidal portion 
by varying the scale of the wavelet 
function. This method produces the power spectrum of the oscillations
in different light curves. We note that the Morlet wavelet suffers from an
``edge effect'' 
that is typical of analyses of time series data. Fortunately, this 
effect is only significant in regions that are
within a cone of influence (COI) that demarks where 
possible periods that are too close to either the
measurement interval or the maximum length of the time series
cannot be convincingly detected. 
The wavelet procedure, its noise filtering capabilities, and the impact 
of COI effects are described in detail in Torrence \& Compo (1998). 

The randomlet software obtains  measurements of the peak power 
in the global wavelet spectrum, 
which is the average peak power over time, and is 
equivalent to a smoothed Fourier power spectrum. 
In principle, the randomization technique compares the average 
values of the actual time series to the peak powers evaluated 
for $n!$ equally-likely permutations of the time series data, 
using the assumption that the $n$ values of 
measured intensities are independent of the $n$ measured times 
if no periodic signal is present. 
The fraction of permutations that provide peak values greater than or 
equal to the original peak 
power of the time series provides the probability ($p$) that
 there is no periodic 
component. Thus the acceptance percentage probability that real 
periodic components are present in the 
data is ($1-p$) $\times$ 100.  To claim a statistically significant 
oscillation period is present
the lowest acceptance probability we allow is 95\%, although it
might be better to call these candidate periodicities.  We consider
really firm detections to have acceptance probabilities of $\ge$ 99\%.

Because computing $n!$ permutations is computationally very expensive, we 
calculate $m = 200$ permutations.  This allows a reliable estimation 
of $p$, as the 95\% confidence interval for the
true value of $p$ is $p \pm 2[p(1-p)/m]^{1/2}$ (Linnell Nemec \& Nemec 1985).
The estimated value of $p$ can have a value of zero, i.e., there is an
almost zero chance that the observed time series oscillations could 
have occurred
by chance. In this case, the 95\% confidence interval can be obtained using the
binomial distribution and is given by 0.0 $< p <$ 0.01, that is, the probability
of a real period having been found is 99$-$100 \% (O'Shea et al. 2001). 

Since the blazar data are also approximately uniformly sampled we can 
use O'Shea's code to produce the power spectra of the IDV on different 
nights. It should be noted that there is no time gap in any of the 
chosen data sequences. We use an average smoothing technique, based on 
low-pass filtering methods, to reduce noise in the original signal before 
searching for possible real periodicities. This
is accomplished through the ``running average" 
process in O'Shea's wavelet tool that smooths the original signal by a 
defined scalar width, taken to be 10. This process uses the ``smooth"
subroutine available with the IDL tool kit which returns a copy of
the array smoothed with a boxcar average of the specified width.
The smoothed signal is then subtracted from the original signal to 
yield the resultant signal used for the wavelet analysis.

\section{Results}

\subsection{Wavelet Analysis Results for IDV Time Scales}

We used high quality optical IDV data of the blazar S5 0716$+$714 taken 
on 20 nights during the period November 26, 1999 to March 23, 2003 from 
Montagni et al.\ (2006). The data were available as magnitudes vs time. 
We converted them into fluxes vs time and then performed the wavelet analysis 
along with a randomization test described in \S 2 on these 20 IDV LCs to search 
for periodicities and IDV time scales for each light curve. 

The LCs, plotted so as to show fractional deviations of intensities from 
the nightly means, are given for several of the nights in the upper-left panels 
of Figs.\ 1$-$4.
The wavelet power transforms of  those LC are shown in the  
middle-left panels of Figs.\ 1$-$4 while the dates the data were taken
 are given in the  
captions. Here we have plotted the wavelet analysis results of 
the three out of the 20 nights during which 
the probability a period (or quasi-period) is really present is 99$-$100\%, 
as well as that for one night during which such a signal was 
much weaker (Fig.\ 4).
In the middle-left panels of these figures
the darkest regions show the most enhanced 
oscillatory power. The cross-hatched areas are the cone of 
influence (COI), the region of 
the power spectrum where edge effects, due to the finite lengths
of the time series, are likely to dominate. Because of this COI 
the maximum possibly detectable periods range between 
10 and 16 ks for different nights.  

In our wavelet 
analyses, we consider time scales that correspond to the 
highest power peaks,
that lie below those COI thresholds and have full-night probability estimates 
above 95\% to be have high significance and thus provide good
period candidates. 
The middle-right panels in the figures show the global wavelet 
power spectra of the LC time 
series from which those periods can be selected.  The width of
those peaks indicate that they are better characterized as
quasi-periods in most cases. We use the term quasi-period here
in the sense of Espaillat et al.\ (2008), i.e., the nominal period is
not crisply defined, as can be seen most clearly in Figs.\ 1 and 3;
however, these signals do not really resemble the QPOs seen in X-ray
binaries (e.g., Remillard \& McClintock 2006), and so, to avoid
confusion with them we will henceforth just use the word ``period''.

Finally, the bottom-left panels show the probabilities of the
presence of two specific frequencies corresponding to the 
first and second highest peaks
shown in the middle-right panels as functions of the time 
after the start of the observations
that night.  The numbers quoted for those periods at the bottom-right
of these figures give the maximum values of the probabilities
and not the global probability level given at the upper-right.
The solid line represents the probability corresponding to
the major power peak of
time series data, while the dotted line corresponds to the 
secondary power peak.  Note that the major power peak shows very
high probabilities for extended intervals during the night, while the
secondary peaks never do so for more than a short period.
We only consider the period corresponding to  the first major power 
peak in our analysis. 

The wavelet analysis results for IDV time scales (or
oscillation periods for a portion of the flux) 
and their probabilities are given in 
columns 5 and 6, respectively, of Table 1
for all 20 nights that satisfied the selection criteria
discussed in \S 3. 
We have only presented in Figs.\ 1--3 the wavelet analysis results for the LCs which
have the highest probability (99$-$100\%) of the existence of statistically
significant time scales.  For the sake of comparison, in  Fig.\ 4 we 
have also presented 
the wavelet analysis result of a LC in which we 
did not find any dominating and statistically significant time scale.
From the intensity wavelet spectrum of Fig.\ 1, we found that the first 
oscillation period of $\sim$3.2 ks dominates over the $\sim$ 50--300 min
span, while 
the second oscillation period of $\sim$5.0 ks dominates over the 
$\sim$ 300 -- 500 
min portion of the time series but not as strongly as did 
the first one. In the global
wavelet spectrum, which is an average of the wavelet spectrum over time
(Torrence \& Compo 1998; O'Shea et al. 2001), the period
corresponding to the maximum power is $\sim$3.2 ks. 
It is also clear from Fig.\ 2 that a period near 4.3 ks is present in the
whole night's time series and always dominates outside the COI.   
In Fig.\ 3 we see that a period of $\sim$1.9 ks is nearly always 
present and is very strong between $\sim$100 and 300 minutes and 
again after 400 minutes that night. 
 
\subsection{Black Hole Mass Estimation}

The mass of the central black hole (BH) in an AGN is probably
the single most important quantity to be known about its central engine. 
The more reliable, or primary, black hole mass estimation methods include
stellar and gas kinematics, reverberation mapping and
megamaser kinematics (e.g.,
Vestergaard 2004). The stellar and gas kinematics
methods require high spatial resolution spectroscopy of the host 
galaxy,  the reverberation mapping method requires detection of 
higher-ionization emission lines from gas close to the BH, 
and megamasers, when present, are
only detectable in essentially edge-on sources. 
Since the optical spectrum of the blazar 0716$+$714 
is a featureless continuum it is not possible to determine the black hole 
mass using the methods that require spectroscopy; and, as
blazars are nearly face-on sources, the megamaser technique is also moot.
Secondary black hole mass estimation methods are either 
approximations to the reverberation mapping approach that still rely on 
the presence of a well measured strong emission line or they employ well-known
empirical relations between the black hole mass and the velocity dispersion 
or mass of the host galaxy's bulge. The marginal detection of the host of 
0716$+$714 (Nilsson et al.\ 2008) means that these approaches to the value  
of its BH mass are also not useful in this case.

Variability time scales can also provide estimates of the mass of an AGN's BH.
An observed variability
doubling time, $\Delta t_{obs}$, provides a crude bound on the mass if
the changing flux arises in the accretion flow close to the BH.
One could then use causality to limit 
the size of the emitting region to
$R \le c \Delta t_{obs} $.
Combining this result with the expectation that the minimum size 
for such an emitting region
is fairly closely related to the gravitational radius of the BH,
$R \ge R_g \equiv GM/c^2$ (e.g., Wiita 1985) one can obtain an estimate for
$M$.  But if the source of the variable
blazar emission is instead from the jet and moving at a velocity
$v$ (e.g., Marscher \& Gear 1985) then
the intrinsic doubling time in the
rest frame of the blazar flow, $\Delta t_{in}$, is given by
$\Delta t_{in} = [\delta / (1+z)] \Delta t_{obs}$, where 
$\delta = [\Gamma (1-\beta {\rm cos}\theta)]^{-1} $ is the Doppler 
factor, with $\beta = v/c$, $\Gamma = (1 - \beta^2)^{-1/2}$ 
and $\theta$ is the viewing angle to the jet axis.
For typical blazars, $\delta \sim 10$, albeit with a substantial
range (e.g., Ghisellini et al.\ 1998).
Earlier measurements of the Lorentz factor for this source
from ejection velocities of radio knots showed a decrease in
the apparent superluminal speeds from $\sim 15c$ to $\sim 5c$ over
the course of a decade 
(Bach et al.\ 2005).  Those changes are  consistent with a constant 
$\Gamma \simeq 12$ but a
decreasing value of $\theta$ from $5^\circ$ to $0^{\circ}.5$ during that
period, leading to a rise in $\delta$ from $\sim 13$ to $\sim 25$
(Montagni et al.\ 2006).  However, it must be stressed that 
even when knot motions can be determined by VLBI measurements, 
obtaining an accurate value of $\delta$ is very difficult, particularly
if the jet has a finite opening angle 
(e.g., Gopal-Krishna, Dhurde \& Wiita 2004).

In the case where an apparent periodic (or nearly periodic) signal is found,
as appears to be the case for 0716$+$714,
then there is some hope to be able to do better with mass estimates.  In these
circumstances, it is probably most reasonable to assume that the period is 
related to the orbital timescale of a blob or flare in the inner portion of 
the accretion disk (e.g., Mangalam \& Wiita 1993). The minimum likely period 
then corresponds to the orbital
period at the inner edge of the accretion disk, which is usually taken to 
be given by the marginally stable orbit, $R_{ms}$, although it is conceivable
that emission can occur from matter at even smaller radii 
(e.g., Abramowicz \& Nobili 1982).  
For a non-rotating (Schwarzschild) BH,
$R_{ms} = 6 GM/c^2 = 6 R_g$, 
while for a maximal
Kerr BH, with angular momentum parameter $a \rightarrow 1$, 
$R_{ms} \rightarrow R_g$. However,
for a more realistic maximum angular momentum parameter of $a = 0.9982$
then $R_{ms} \simeq 1.2 R_g$ (e.g., Espaillat et al.\ 2008).

The angular velocity of co-rotating matter orbiting a BH, as measured 
by an inertial observer at infinity is given by (e.g., Lightman et al.\ 1975)
\begin{equation}
\Omega = \frac{ M^{1/2}} {r^{3/2} + a M^{1/2}}~,
\end{equation}  
where geometrical units,  $G = c = 1$, have been used,
 and $r = R/R_g$.
This leads to an expression for the BH mass in terms of the observed
period, $P$, in seconds,
\begin{equation}
\frac {M} {M_{\odot}} = {\frac {3.23 \times 10^4 ~ P} {(r^{3/2} + a)(1+z)}}.
\end{equation}
The nominal masses obtained in this fashion for a Schwarzschild
BH (with $r = 6$ and $a = 0$) are column 7 of Table 1,
and those obtained for a maximal Kerr BH
(with $r = 1.2$ and $a = 0.9982$) are in column 8.
If the periodic perturbations in the inner part of the disk are
advected into the jet and the observed emission comes from a relativistic
flow directly affected by those perturbations then the value of $M$ 
would require the results obtained from
Eq.\ (4) to be multiplied by $\delta$, so these mass estimates are
really lower bounds.

\subsection{Correlation between Flux and IDV Time Scale}

In our sample of 20 clearly variable IDV LCs, 16 were measured in the
R band but 4  were taken using a V filter. Those V band measurements
have to be converted to R in order to investigate any correlation 
between the flux level and quasi-periodic variation.  Sagar et al.\ 
(1999) reported an average V$-$R color of this BL Lac to be
 $\sim$ 0.4 mag.\ in their one month long BVRI optical monitoring campaign
in 1994.  
We adopted this value of V$-$R and converted the  V magnitudes
 of the four  LCs into R magnitudes and then calculated average nightly
fluxes in the R band for them. 
We also computed the average fluxes of the 16 nights where the LCs are
already in the R band. 

In Fig.\ 5 we plot the average nightly flux vs highest probability
IDV time scales ($P$), noting however, that some of the timescales have rather
low probabilities of being real. 
We see that there is no correlation between the average flux and the IDV 
time scales of the blazar, regardless of whether all 20 points or the
5 with the highest probabilities are considered.  It is worth noting
that the most confident periods are only detected on nights when the
total flux is  high.  This is reasonable if larger mean fluxes correspond
to times when the periodic variable component is stronger.

These  20 nights  were spread over more than 3 years (November 26, 1999 
to March 23, 2003). Since the source has shown long-term 
quasi-periodicity on the time scale of 
$3.0 \pm 0.3$ years (Gupta et al.\ 2008),  this blazar has gone
through most, if not all, of its (so-far observed) long-term states 
during the period in which the data for the present work was obtained. 
 It is generally 
believed that in the pre-outburst, outburst and post-outburst states, even
the optical emission can be 
attributed to the shock moving down the inhomogeneous medium in the jet
(e.g., Marscher et al.\ 1992). On the other hand, 
in the low state of a blazar,  IDV  arising 
from instabilities
 in the accretion disk could more easily be detected
(e.g., Mangalam \& Wiita 1993).

In blazars, the radiation emitted by the  plasma in the jet, which has
bulk relativistic motion and is oriented at small viewing angles, will
be affected by relativistic beaming, which in 
turn implies a shortening of the observed timescales by a factor 
of $\delta^{-1}$. 
A correlation between average flux and IDV 
time scales should  be expected if the variability arises from
changes in the
velocity of, and/or viewing angle to, the emitting region.
We do not see such a correlation, which implies that optical emission 
from this blazar is not governed by this single mechanism and
thus that more than one source of the radiation is probably present 
at these times. 

\section{Discussion and Conclusion}

There are two major classes of models for AGN variability, those involving
shocks-in-jets and those invoking instabilities on or above accretion disks.
The former is expected to dominate in blazars and the latter is most important
when jets are absent or weak (e.g., Wagner \& Witzel 1995; 
Urry \& Padovani 1995; Mangalam \& Wiita 1993). In another jet-based 
variant it is argued that
for low-luminosity AGNs, the accretion disk is radiatively inefficient 
and any small amplitude variation, even in the low-state of the source, will be 
due to a weak jet (e.g., Chiaberge et al.\ 1999, 2006; Capetti et al.\ 2007, 
and references therein). This variant may be able  to  explain 
optical IDV in blazars. The detection of quasi-periodicity or periodicity 
on IDV time scales can be most easily explained by the presence of a single 
dominating hot-spot on the accretion disk (e.g., Mangalam \& Wiita 1993; 
Chakrabarti \& Wiita 
1993) or perhaps by pulsational modes in the disk 
(e.g., Espaillat et al.\ 2008). 

Together, the difficulties in obtaining lengthy, high quality, and essentially
evenly spaced data for ground based observations of AGN, 
and the weaknesses of the standard analysis tools
(periodograms and Fourier transforms) under these circumstances, has
meant that
performing good searches for periods or quasi-periods in blazars
has been difficult
until recently.  So it is not surprising that 
there have been few strong claims of quasi-periodic variations in blazar 
LCs. Carini et al.\ (1992) noted a weak peak in the power spectrum 
of optical IDV observations of OJ 287 at 
$\approx$ 32 minutes, but it  was not statistically 
significant. In the same blazar, 
Carrasco et al.\ (1985) earlier claimed periodic variations on timescales 
of tens of minutes but they were not convincing.
Earlier excellent optical data of the blazar 0716$+$714 
has shown possible quasi-periodicity  on the timescales of $\sim$ 1 day, 
4 days and 7 days, and some of these were in apparent synchrony with
radio variations, but the time series were too short to provide conclusive
evidence (Wagner 1992; Heidt \& Wagner 1996).
Very recently, using X-ray data for the blazar 3C 273, 
Espaillat et al.\ (2008) have
reported quasi-periodicity on a timescale of 3.3 ks.  If these
fluctuations represent orbital periods they would imply central BH masses 
for 3C 273 some 10 to 100 times smaller than independent determinations
have found, so Espaillat et al.\ (2008) favor the hypothesis that the
variations they have seen arise from higher-order modes within the disk.  
To our knowledge, no other members of the 
blazar class have shown significant harmonic components 
in the IDV or short-time scale LCs.   

Here we have used a wavelet plus randomization technique 
(Linnell Nemec \& Nemec 1985; O'Shea et al.\ 2001) to search for 
optical IDV time scales of the blazar S5 0716$+$714. 
We selected 20 nights of IDV data that were characterized by long strings of 
approximately uniformly sampled data with high IDV signals
and low noise from an initial database of 102 nights of
observations by Montagni et al.\ (2006).  

We found high probabilities ($ \ge 95$\%) of (at least quasi-) periodic 
oscillations in 
eight of these IDV LCs and very high probabilities ($ \ge 99$\%) 
for five of them.
 This is the first evidence of detection of periodic components
in blazar optical IDV LCs. We found quasi-periodic IDV time scales between 
$\sim$ 25 and $\sim$ 73 minutes.

These variability timescales lead to nominal BH  masses ranging from 
2.47 -- 7.35 $\times$ 10$^{6}$ M$_{\odot}$ 
assuming the period
arises from a Schwarzschild BH, while for a rapidly rotating Kerr BH
these mass estimates rise to 1.57 -- 
4.67 $\times$ 10$^{7}$ M$_{\odot}$.  
However, if these variations arise from internal disk modes then the
corresponding BH mass can be substantially larger, by factors of
$\sim$10 to over 100.   And if they do emerge 
from an inner portion of the  jet where they are induced by 
disk fluctuations, the masses 
would be increased by a factor of $\delta \sim 20$; however, if
they emerge from jets at large distances from the SMBH, then clearly
no information on the BH mass can be extracted from these timescales.

  We did not find 
any correlation between average flux and IDV time scales,
which implies that optical emission from the 
blazar is probably not governed by a single mechanism.
Our results appear to be most consistent with models in which accretion
disk variability plays a role in the optical emission of blazars. 
Nonetheless, it is possible that emission from turbulet jet, 
which is also precessing or swinging (e.g., Camenzind \& Krockenberger 1992; 
Gopal-Krishna \& Wiita 1992) could also explain these observations.

\acknowledgments

We are extremely grateful to Dr.\ E.\ O'Shea for kindly providing 
and allowing us to use his wavelet code for the present work. 
We gratefully acknowledge Prof.\ M.\ H.\ P.\ M.\ van Putten
for carefully reading an earlier version of the manuscript and thank
the anonymous referee for suggestions that have improved the presentation. 
This research made use of the SIMBAD database, operated at 
CDS, Strasbourg, France.  PJW's work is 
supported in part by a subcontract to GSU from NSF grant AST 05-07529 
to the University of Washington.

\clearpage

\begin{deluxetable}{cccccccc}
\tabletypesize{\scriptsize}
\tablecaption{Wavelet analyses for IDV of the blazar S5 0716$+$714 
\label{tbl-1}}
\tablewidth{0pt}
\tablehead{
\colhead{Date} & \colhead{Band} & \colhead{Duration} & \colhead{Amplitude} &
\colhead{Time Scale} & \colhead{Probability} 
& \colhead{BH Mass (Sch.)} & \colhead{BH Mass (Kerr)} \\
\colhead{(dd.mm.yyyy)} &          & \colhead{(hours)}   & \colhead{(\%)} &
\colhead{(seconds)} & \colhead{(\%)}
& \colhead{(10$^6$ M$_{\odot}$)}   & \colhead{(10$^7$ M$_{\odot}$)}             
}
\startdata
26.11.1999 & V     & 11.34 & 36   & 4295 & 95.0 & 7.21 & 4.58 \\
12.01.2000 & V     & 11.40 & 32   & 3205 & 82.0 & 5.38 & 3.42 \\
25.01.2000 & V     & 10.74 & 32   & 3301 & 92.0 & 5.54 & 3.52 \\
27.10.2000 & V     & ~7.72 & 19   & 1473 & 99.0 & 2.47 & 1.57 \\
14.02.2001 & R     & 10.55 & 11   & 1978 & 92.5 & 3.32 & 2.11 \\
26.02.2001 & R     & 10.00 & 29   & 2050 & 96.0 & 3.44 & 2.19 \\
03.11.2001 & R     & 10.04 & 15   & 1670 & 90.5 & 2.80 & 1.78 \\
01.02.2002 & R     & ~8.19 & 12   & 3039 & 91.5 & 5.10 & 3.24 \\
13.03.2002 & R     & ~8.77 & 10   & 3530 & 30.0 & 5.92 & 3.76 \\
15.03.2002 & R     & ~9.23 & 14   & 3004 & 58.5 & 5.04 & 3.20 \\
20.03.2002 & R     & ~9.32 & 10   & 2415 & 93.5 & 4.05 & 2.57 \\
25.03.2002 & R     & ~9.36 & 18   & 3226 & 99$-$100 & 5.41 & 3.44 \\
22.04.2002 & R     & ~7.68 & 22   & 4298 & 99$-$100 & 7.21 & 4.58 \\
29.12.2002 & R     & 12.25 & 53   & 1917 & 95.0 & 3.22 & 2.04 \\
18.02.2003 & R     & 10.92 & 21   & 3759 & 94.0 & 6.31 & 4.01 \\
04.03.2003 & R     & ~9.40 & 39   & 2041 & 83.0 & 3.42 & 2.18  \\
18.03.2003 & R     & ~8.63 & 15   & 1890 & 99$-$100 & 3.17 & 2.01  \\
23.03.2003 & R     & ~9.16 & 24   & 4380 & 99.5 & 7.35  & 4.67 \\
\enddata


%

\end{deluxetable}

\clearpage

\begin{figure}
\plotone{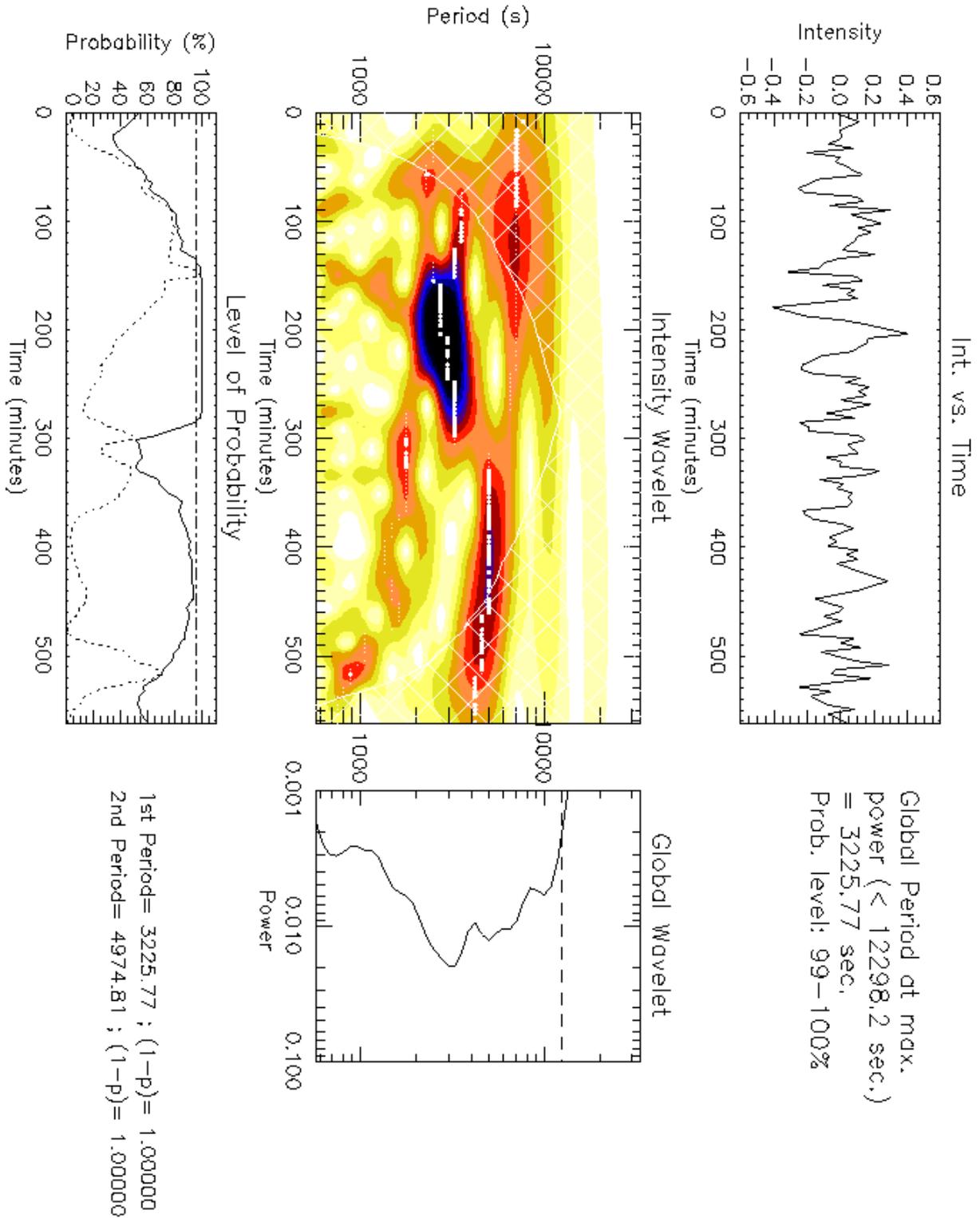}
\caption{Wavelet analysis for the light curve of the blazar S5 0716$+$714 for 
the date March 25, 2002. The top panel shows the variation of fluxes computed 
from the data of Montagni et al.\ (2006) with the overall most significant 
period described at the top right. The wavelet power spectrum is given in the 
middle panels, with the shortest scales at the bottom. The probabilities 
of the two strongest peaks being real at different times are plotted in the 
bottom left panel with the maximum probabilities given at the bottom right.  
See text for details. A color version of this figure will appear in the 
electronic version of the Journal.} 
\end{figure}

\clearpage

\begin{figure}
\plotone{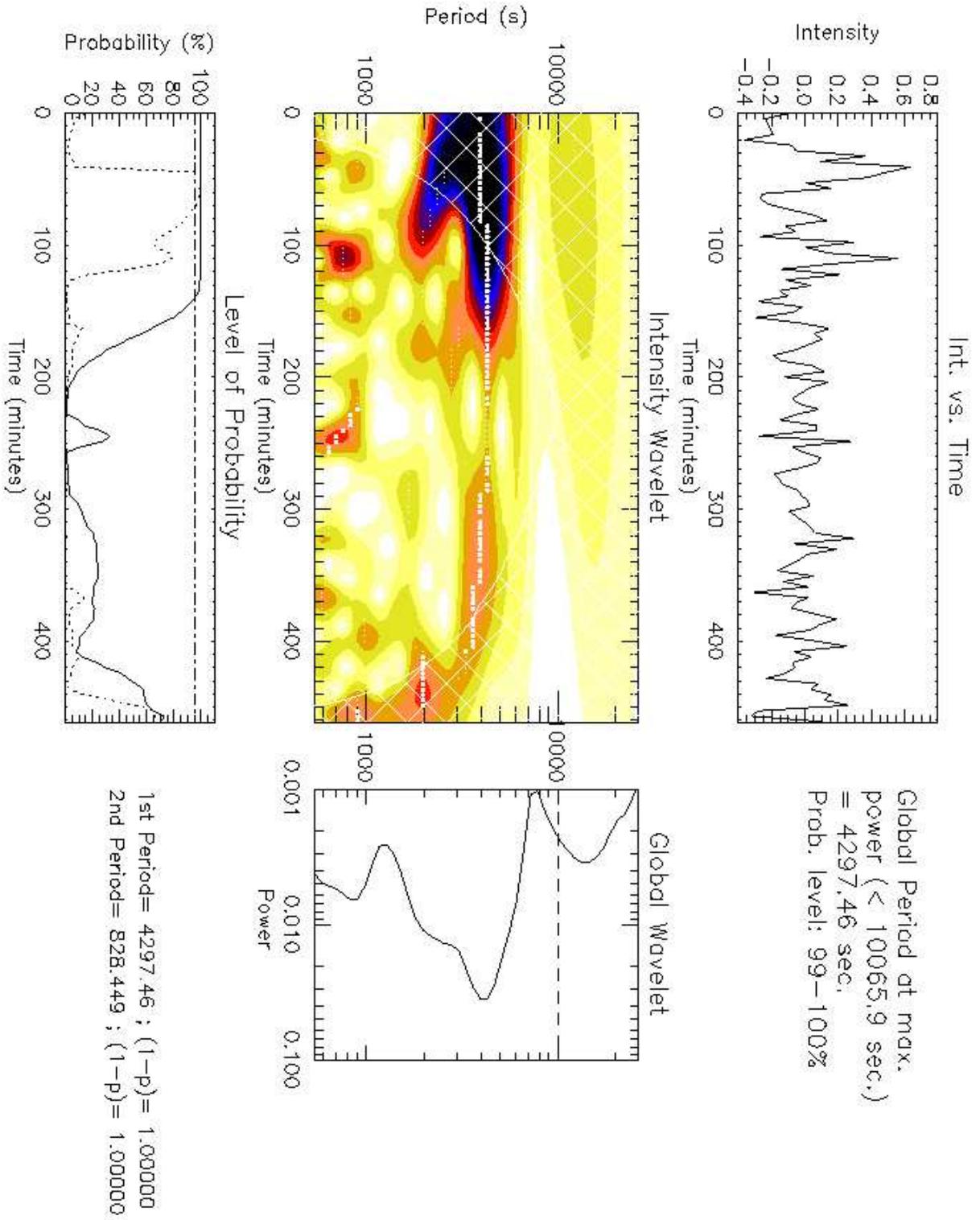}
\caption{As in Fig.\ 1 for April 22, 2002.}
\end{figure}

\clearpage

\begin{figure}
\plotone{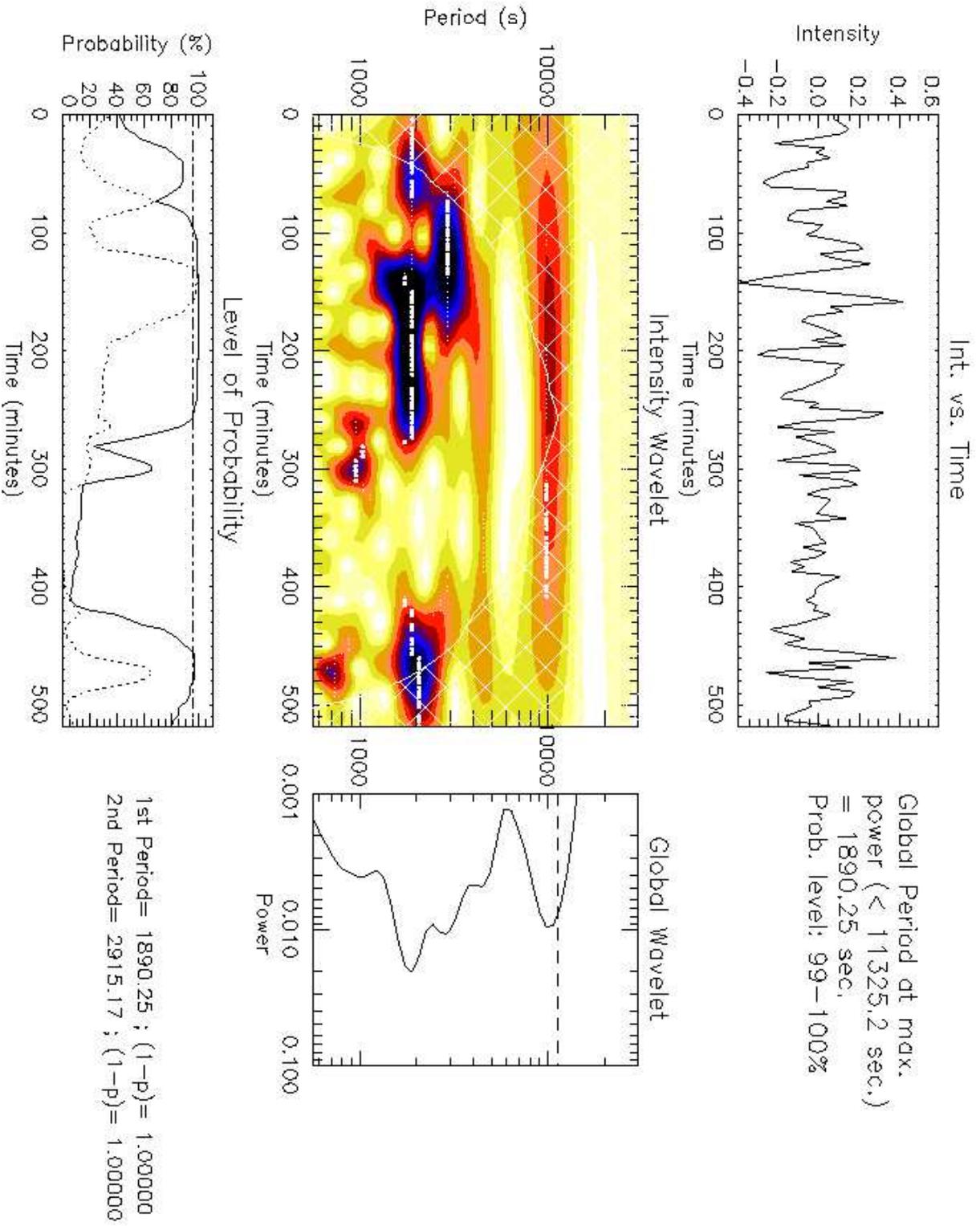}
\caption{As in Fig.\ 1 for March 18, 2003.}
\end{figure}

\clearpage

\begin{figure}
\plotone{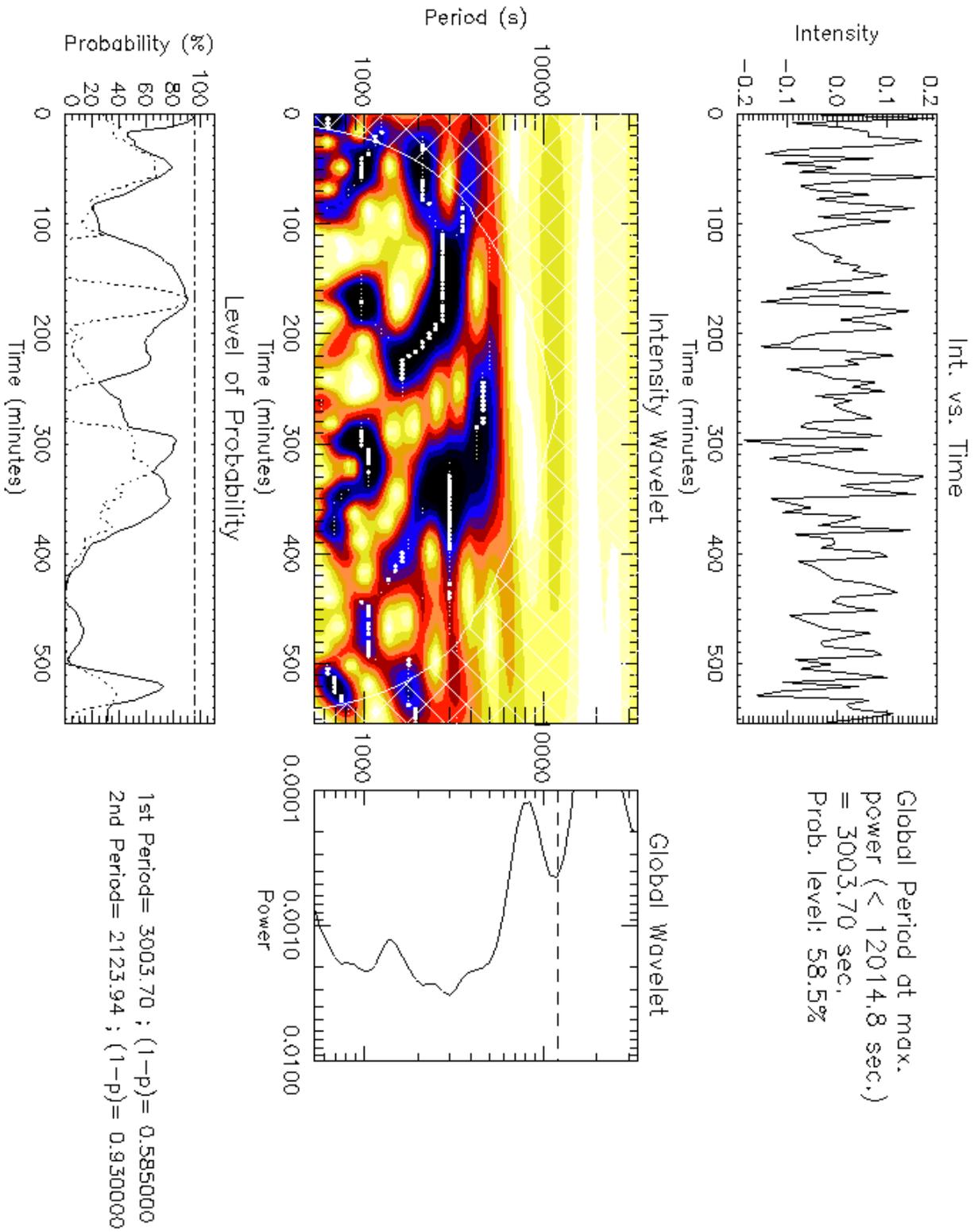}
\caption{As in Fig.\ 1 for March 15, 2002.}
\end{figure}

\clearpage

\begin{figure}
\plotone{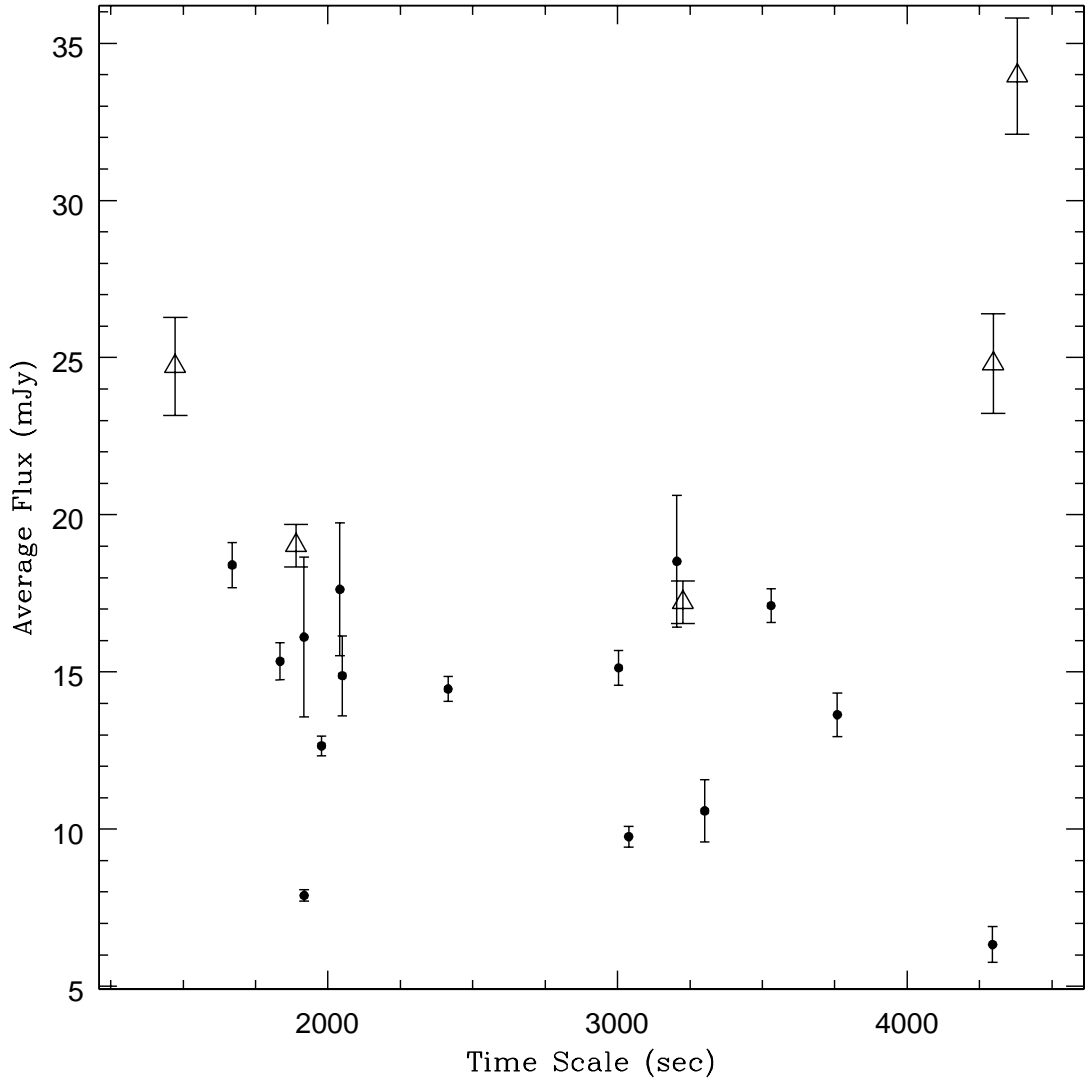}
\caption{Average nightly R band flux vs IDV timescale for the sample
of 20 LCs given in Table 1. Open triangles indicate that
the probability of detection of a periodic component in the variability is
$\geq$ 99\%.}
\end{figure}

\end{document}